\documentclass[preprint,showpacs,amsmath,amssymb]{revtex4}

\usepackage{graphicx}
\usepackage{dcolumn}
\usepackage{bm}
\usepackage{multirow}
\usepackage{amsmath}
\usepackage[version=3]{mhchem}
\usepackage{float}

\begin{document}
\title{Nature of Ionic Diffusion and Electronic Transfer in Eldfellite Na$_x$Fe(SO$_4$)$_2$}

\author{Chol-Jun Yu$^{1*}$, Song-Hyok Choe$^1$, Gum-Chol Ri$^1$, Sung-Chol Kim$^1$, Hyok-Su Ryo$^2$, and Yong-Jin Kim$^2$}
\affiliation{$^1$Department of Computational Materials Design (CMD), Faculty of Materials Science, and\\
$^2$Faculty of Physics, Kim Il Sung University, Ryongnam-Dong, Taesong District, Pyongyang, DPR Korea}
\date{\today}
\begin{abstract}
Discovering new electrodes for sodium-ion battery requires clear understanding of the material process during battery operation. Using first-principles calculations, we identify mechanisms of ionic diffusion and electronic transfer in newly developed cathode material, eldfellite Na$_x$Fe(SO$_4$)$_2$, reproducing the electrochemical properties in good agreement with experiment. The inserted sodium atom is suggested to diffuse along the two-dimensional pathway with preceding movement of the host sodium atom, and the activation energy is calculated to be reasonable for fast insertion. We calculate the electronic properties, showing the band insulating at low composition of inserted sodium, for which the electron polaron formation and hoping are also suggested. Our results may contribute to opening a new way of developing innovative cathode materials based on iron and sulfate ion.
\end{abstract}
%82.47.Aa: Batteries, lithium ion
%71.20.Tx: Intercalation compounds
%66.30.Pa: Diffusion in nanoscale solids
%71.38.-k: Polarons in electronic structure of solids
%71.15.Mb: Density-functional theory: condensed matter
\maketitle
The scarcity and price rise of lithium due to its massive consumption for lithium-ion batteries (LIBs) during the past two decades~\cite{Tarascon10,Goodenough10} prompts to develop another ionic battery based on sodium that has abundant natural resource and much lower cost~\cite{Chevrier,Palomares,Slater,Yabuuchi14,Sawicki,Wang15}. When compared to Li, however, Na has two intrinsic drawbacks: Its lower ionization potential leads to lower operating voltage and lower power density, and its larger ionic radius can cause the slow ionic diffusion and larger volume change at the electrodes during charge-discharge process. These difficulties become even more pronounced in the cathode. Therefore, extensive seeking for unique crystalline structures with an open framework or large insertion channels to facilitate the intercalation of \ce{Na+} ion with high voltage over 3 V for the cathode is vital to the development of commercially viable sodium-ion battery (SIB).

The latest discoveries of iron-based polyanionic sulfate cathodes~\cite{ReynaudPhd,Barpanda14, Singh,Reynaud14,Mason,Meng} raise hope to replace nowadays ubiquitous LIBs with low-cost SIBs for large-scale grid energy storage and electric vehicles~\cite{Armand,Dunn}. The interest comes not only from the highest electrode potentials with stable reversible capacity among polyanionic cathodes discovered so far, thanks to a high electronegativity of sulfate ion \ce{SO4^2-}, but also from the earth-abundance of inexpensive, non-toxic transition metal iron resource. In fact, \ce{Na2Fe2(SO4)3} shows an average voltage of 3.8 V with a reversible capacity over 100 mAh/g for \ce{Fe^2+}/\ce{Fe^3+} redox couple versus Na metal anode~\cite{Barpanda14}. Besides, several cathodes based on Fe and sulfate ion have also been identified from the experiments, such as \ce{Na2Fe(SO4)2}~\cite{Reynaud14}, \ce{Fe2(SO4)3}~\cite{Mason} and \ce{Na2Fe(SO4)2\cdot}2\ce{H2O}~\cite{Meng}.

Here we focus on a newly synthesized material, eldfellite \ce{NaFe(SO4)2}, which was found to be a potential SIB cathode with an average voltage of $\sim$3.0 V and a capacity near 80 mAh/g for a relatively long life~\cite{Singh}. Moreover, there are further rooms for this material such as magnetic analogues with manganese or nickel to have higher power density, and reducing the number of sulfate linkages to increase the capacity. Indeed, these can be realized only based on clear understanding of the fundamental physics and material properties of eldfellite, but not yet explored.

In this Letter, we present a first-principles study of sodium ion diffusion and electron transfer upon sodium insertion into eldfellite \ce{NaFe(SO4)2}, {\it i.e.,} \ce{Na_{\it x}Fe(SO4)2} ($0.75\leq x\leq1.75$). By using a Hubbard $U$ augmented density functional theory (DFT+$U$) approach~\cite{Cococcioni}, we first calculate the electrochemical properties including volume change, operating voltage and binding energy of intercalated sodium ion, which are in good agreement with available experimental data. Then, possible migration paths for sodium ion diffusion are predicted by estimating bond valence sum (BVS)~\cite{Brown_BVS,Adam_BVS2} and the activation energy is calculated by applying the climbing image nudged elastic band (NEB) method~\cite{NEB}. Based on electronic structure calculations, we get a valuable insight that the electronic transfer in Na$_x$Fe(SO$_4$)$_2$ is polaron-like formed by Jahn-Teller distortion of lattice.

All calculations in this work were performed with the pseudopotential plane-wave code Quantum ESPRESSO (version 5.3)~\cite{QE}. The ultrasoft pseudopotentials~\footnote{We used the pseudopotentials Na.pbe-sp-van\_ak.UPF, Fe.pbe-sp-van\_ak.UPF, S.pbe-van\_bm.UPF, O.pbe-van\_ak.UPF, and Li.pbe-s-van\_ak.UPF from http://www.quantum-espresso.org.} were used to describe the ion-electron interaction while the Perdew-Burke-Ernzerhof variant of the generalized gradient approximation (GGA)~\cite{PBE} was used for the exchange-correlation interaction between electrons. Both ferromagnetic and antiferromagnetic (AFM) spin polarization were taken into consideration and only the results with AFM order were presented. The on-site effective $U$ parameter ($U_\text{eff}=U-J$) for the localized Fe $3d$ states was set to be 4.0 eV~\cite{ZhouB1,Cococcioni}. Structural relaxations of (2$\times$2$\times$1) supercells including 4 formula units (f.u.) were performed with a 60 Ry plane-wave cutoff energy and a (2$\times$2$\times$4) $k$-point mesh, which guarantee an accuracy of 5 meV/f.u. for total energy. Methfessel-Paxton smearing approach with a 0.2 Ry gaussian spreading factor was used~\cite{mpsmearing}. All atoms are relaxed until the forces converge to 5.0$\times10^{-4}$ Ry/Bohr. See Supplemental Material for more details regarding calculations and analysis.~\footnote{See Supplemental Material at URL for more details regarding calculations and analysis.}

Before proceeding to study the nature of ionic mobility and electronic conductivity in \ce{Na_{\it x}Fe(SO4)2}, we start by systematically analyzing the crystalline lattice change and electrode properties for these systems. \ce{NaFe(SO4)2} was identified to crystallize in a layered structure with monoclinic space group $C2/m$~\cite{Zunic}. In Table~\ref{tab_lattice}, we summarize the optimized lattice constants and volumes of (2$\times$2$\times$1) supercells of \ce{Na_{\it x}Fe(SO4)2} as increasing Na content $x$ from 0.75 to 1.75 with an interval of 0.25. The positions of inserted Na atoms were first estimated by analyzing the difference of bond valence sum from the ideal value ($\Delta$BVS)~\cite{Brown_BVS,Adam_BVS2}, and then fixed by structural optimization. Note that for the case of \ce{NaFe(SO4)2}, our work slightly overestimates the experimental lattice constants ($a=9.520$ \AA, $c=7.115$ \AA, $\alpha=91.63^\circ$, $\gamma=65.91^\circ$)~\cite{Zunic} by $\sim$3.3\%, which is reasonable compared to the typical GGA+$U$ calculations~\cite{ZhouB1}.
\begin{table}[!t]
\caption{\label{tab_lattice} Lattice constants and volume in (2$\times$2$\times$1) supercells of \ce{Na_{\it x}Fe(SO4)2}, binding energy ($E_b$) of inserted Na atom, and electrode voltage, as calculated using $U=4$ eV.}
\begin{ruledtabular}
\begin{tabular}{cccccccc}
 & $a$  & $c$ & $\alpha$  & $\gamma$ & Volume & $E_b$  & Voltage \\
 \cline{2-3} \cline{4-5} 
$x$ & \multicolumn{2}{c}{(\AA)} & \multicolumn{2}{c}{($^\circ$)} & (\AA$^3$) & (eV) & (V) \\
\hline
0.75 & 9.888 & 7.329 & 91.71 & 64.88 & 648.212 & $-5.45$ & 4.22 \\
1.00 & 9.834 & 7.283 & 91.70 & 64.99 & 638.892 &       &      \\
1.25 & 9.886 & 7.311 & 91.71 & 64.88 & 646.587 & $-4.25$ & 3.07 \\
1.50 & 9.959 & 7.285 & 91.73 & 64.72 & 651.462 & $-4.16$ & 3.00 \\
1.75 & 9.993 & 7.228 & 91.74 & 64.64 & 652.844 & $-4.08$ & 2.90 \\
\end{tabular}
\end{ruledtabular}
\end{table}
\normalsize

Regarding the volume change upon Na desertion from \ce{NaFe(SO4)2} ({\it i.e.}, for \ce{Na_{0.75}Fe(SO4)2}), the lattice volume is a little expanded with a relative volume expansion rate ($r_\text{vol}=(V_x-V_{1.0})/V_{1.0}\times100$\%) of 0.7\%, indicating a weakening of electrostatic interaction between layers composed of \ce{FeO6} and distorted \ce{NaO6} octahedra due to the depletion of \ce{Na+} ions. Upon Na insertion into \ce{NaFe(SO4)2}, we also observe the volume expansion with gradual increase of interlayer distance $a$ while decrease of lattice constant $c$ going from $x=1.25$ to 1.75. This might be due to a strengthening of repulsion between layer cations (\ce{Fe^{3+}}/\ce{Fe^{2+}} and the host \ce{Na+} ions) and inserted \ce{Na+} ions in the $a$ (and $b$) direction while attraction between \ce{SO4^{2-}} anions and inserted \ce{Na+} ions in the $c$ direction. However, the relative expansion rates ($r_\text{vol}$) are remarkably small as 1.2, 1.8 and 2.5\% for $x=1.25$, 1.5 and 1.75, compared to almost 50\% in the layered metal oxides~\cite{Delmas,Kim}, and also much smaller than the appropriate value of 5\% for polyanionic electrode suggested by Tripathi {\it et al}~\cite{Tripathi}. Therefore, it can be believed that this material has no such a problem as capacity reduction by irreversible structural change induced by insertion-desertion of \ce{Na+} ion during charge-discharge process, which is agreed well with the experimental result reporting that there is little capacity loss after 80 cycles at a 0.2C rate and the capacity is recovered by almost 100\% when decreasing the rate from 2C to 0.1C~\cite{Singh}.

\begin{figure}[!t]
\includegraphics[clip=true,scale=0.5]{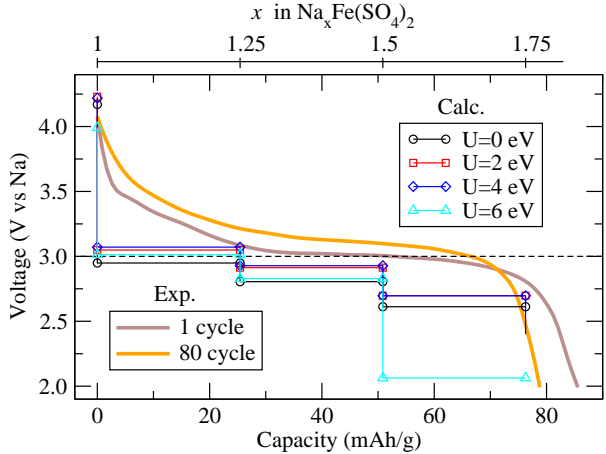} 
\caption{\label{fig_vol}(color online) Electrode voltage versus Na metal anode as a function of capacity (and $x$ in \ce{Na_{\it x}Fe(SO4)2}). Different $U$ values (0, 2, 4, 6 eV) for Fe in GGA+$U$ method are tested. The experimental results of the 1st and 80th cycles at 0.2C rate are shown by solid brown and orange lines, together with the average voltage of $\sim$3 V~\cite{Singh}.}
\end{figure}

As can be appreciated in Table~\ref{tab_lattice}, binding energies of inserted Na atom for all the configurations are negative, indicating a thermodynamically favorable chemical interaction between Na atom and \ce{NaFe(SO4)2} compound. The average discharge voltages versus sodium metal anode from the state of $x=1.0$ were calculated to be 3.07, 3.00, and 2.90 V to the states of $x=1.25$ (25.4 mAh/g), 1.50 (50.9 mAh/g), and 1.75 (76.3 mAh/g) over Fe$^{2+}$/Fe$^{3+}$ redox couple, which are in good agreement with the experimentally identified charge-discharge profiles~\cite{Singh,ReynaudPhd} and average voltage of $\sim$3.0 V at a 0.1C rate with a capacity near 80 mAh/g~\cite{Singh}. Fig.~\ref{fig_vol} shows the calculated step voltages with the experimental result. Such a small volume change and reasonably high electrode voltage indicate a solid feasibility of using \ce{Na_{\it x}Fe(SO4)2} as a promising cathode for SIBs.

We then turn to the main question of how \ce{Na+} ions diffuse and simultaneously charge carriers (electron and/or hole) are transferred inside the electrode during charge-discharge process. These are of significant interest when elucidating the mechanism of battery operation as the sodium ions shuttle through electrolyte while the electrons travel through external circuit between cathode and anode, passing through the electrodes. Moreover, the details of such phenomena are often difficult to extract from experiment alone, especially for new polyanionic framework compounds~\cite{Islam}.

\begin{figure}[!t]
\begin{center}
\includegraphics[clip=true,scale=0.19]{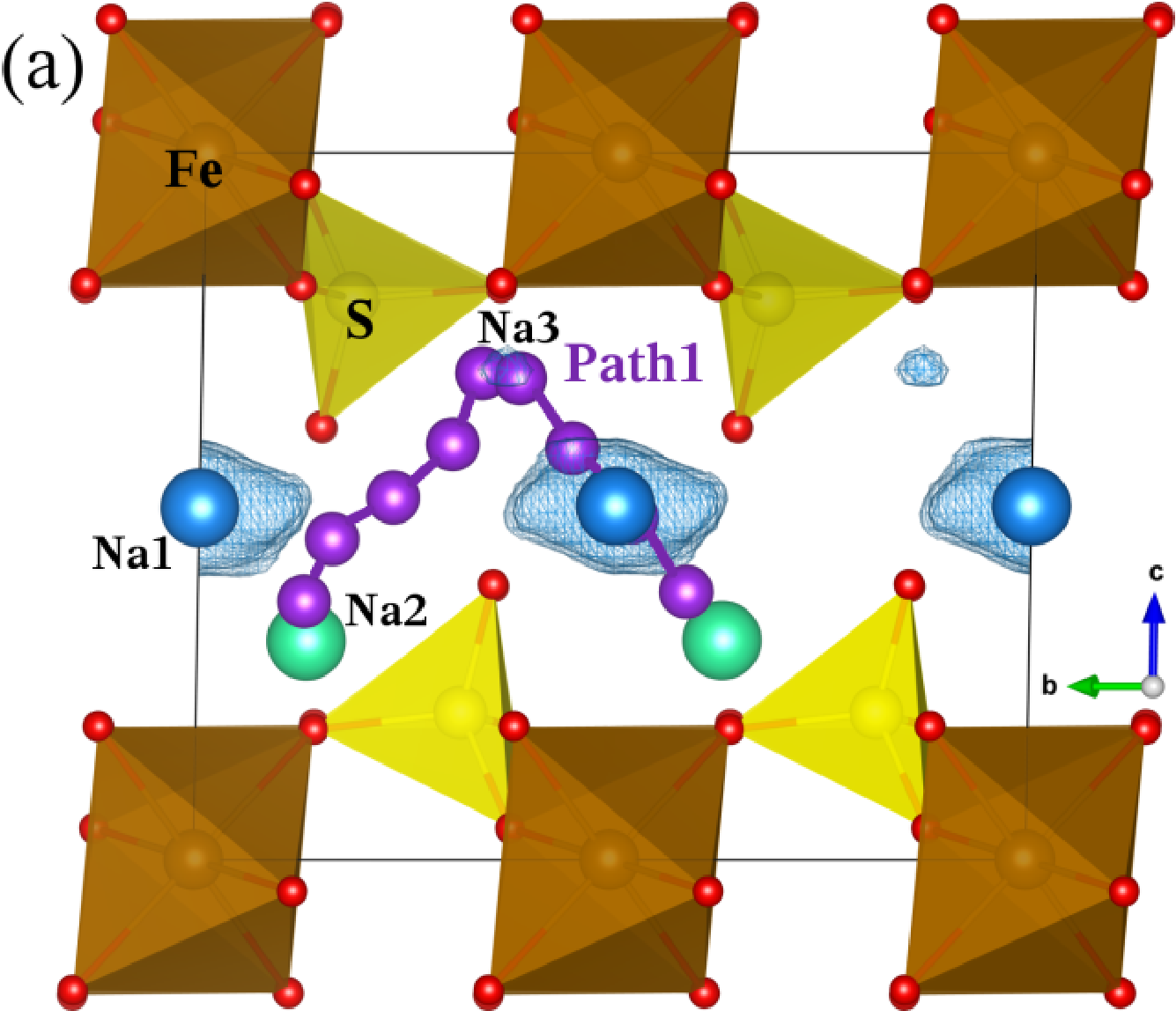}
\includegraphics[clip=true,scale=0.19]{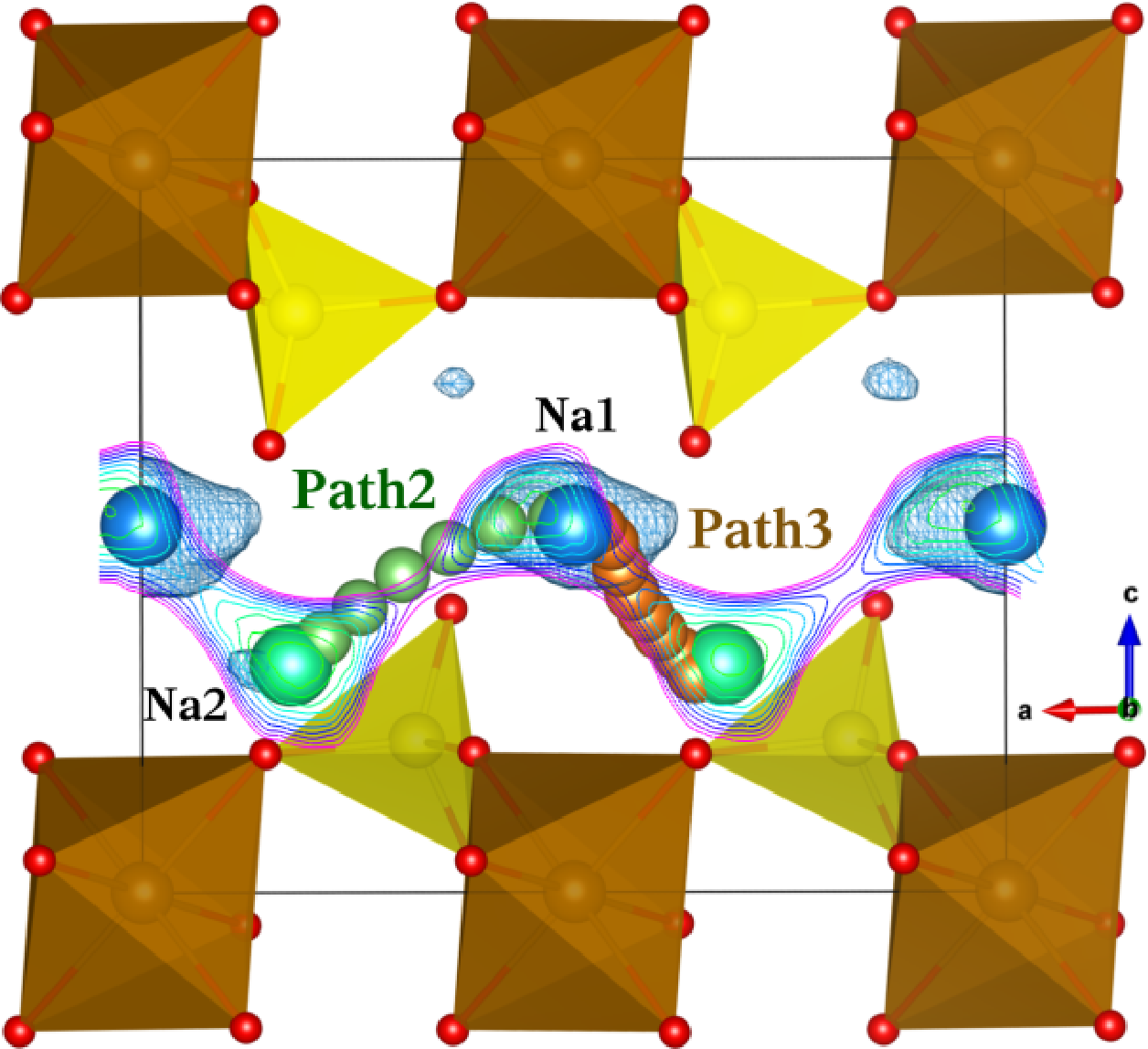}
\includegraphics[clip=true,scale=0.16]{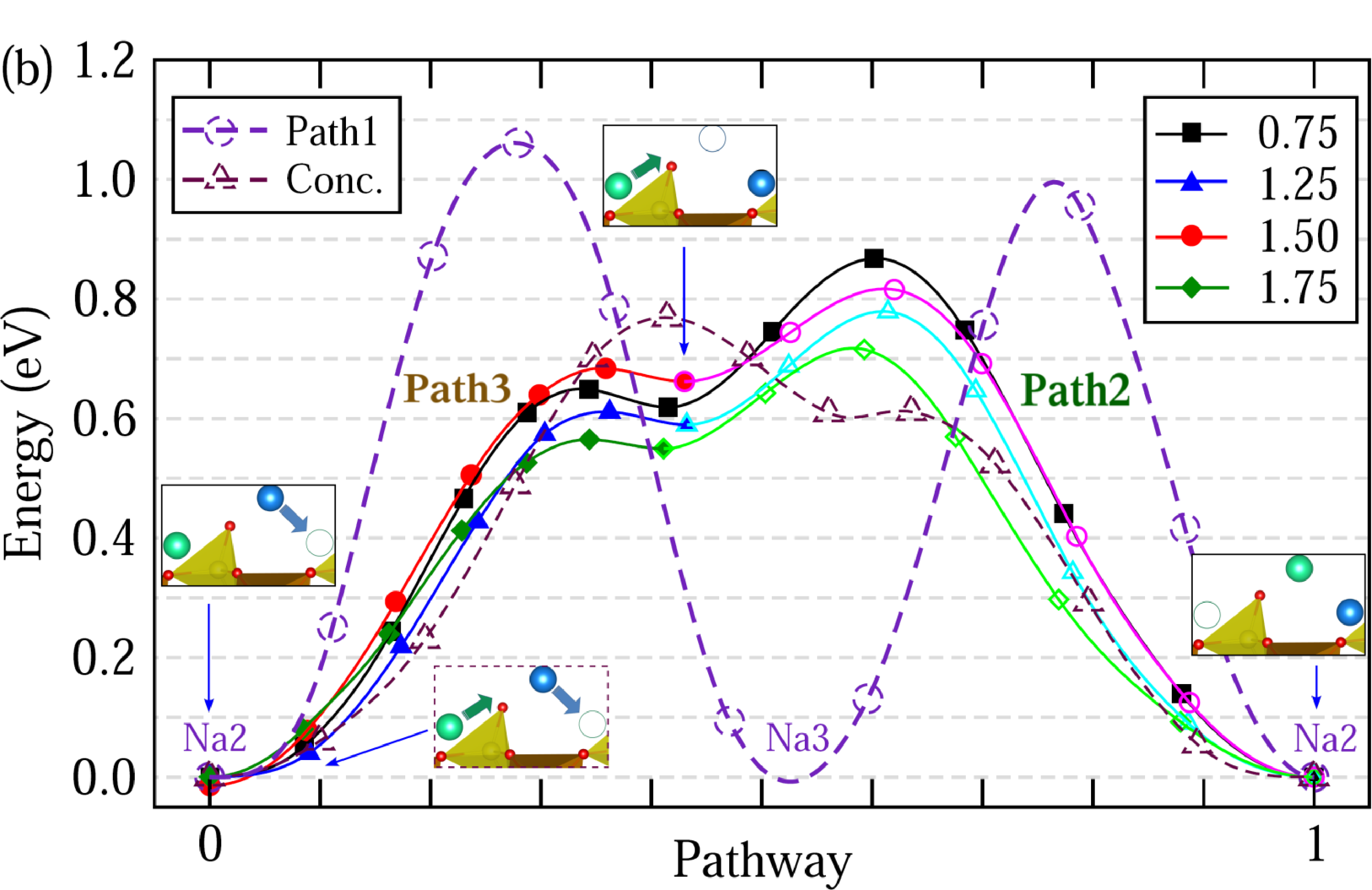}
\caption{\label{fig_mig}(color online) (a) Polyhedral view of $(2\times2\times1)$ supercell of bulk \ce{Na_{1.25}Fe(SO4)2} with an isosurface of $\Delta$BVS (value = 0.3) and migration pathways, where big blue, big green and small red balls represent the host sodium, inserted sodium, and oxygen atoms, respectively. In the right panel a contour of $\Delta$BVS (value = [0.3, 1.3]) is plotted. (b) The corresponding activation energies calculated by NEB method, where Conc. means concurrent move.}
\end{center}
\end{figure}

To investigate the sodium ion diffusion, we estimated possible migration pathways and calculated activation barriers for the diffusion in \ce{Na_{\it x}Fe(SO4)2} by applying the NEB method~\cite{NEB} in combination with $\Delta$BVS analysis~\cite{Brown_BVS,Adam_BVS2}. Fig.~\ref{fig_mig}(a) shows a polyhedral view of the $(2\times2\times1)$ supercell of \ce{Na_{1.25}Fe(SO4)2} with a map of isosurface of $\Delta$BVS and migration pathways for inserted sodium atoms. Here $\Delta$BVS are plotted at the value of 0.3, and the identified positions are consistent with the refined Na positions, which are Na1 site for the host sodium atoms (represented by big blue balls) and Na2 or equivalently Na3 sites for the inserted sodium atoms (by big green balls).

The migration pathways for the inserted sodium atoms were also predicted by the $\Delta$BVS map and proved to be consistent with those fixed by the NEB calculation. First, we consider possible pathways for inserted sodium atom migration using the \ce{Na_{1.25}Fe(SO4)2} model. As clearly shown by $\Delta$BVS map in Fig.~\ref{fig_mig}(a), two pathways are plausible for the inserted sodium atom at the Na2 site to go to the neighboring vacant Na2 site; one way is through the vacant Na3 site, and another way is through the Na1 site that is occupied by the host sodium atom. Since the former case can be a single pathway, we denote as path1. However, the latter case consists of two pathways, denoted by path2 and path3, supposing that the host sodium atom can move also. In this suppose, we can think of two modes; (i) the host sodium atom at the Na1 site and the inserted sodium atom at the Na2 site can move concurrently (concurrent mode), and (ii) the host atom moves first to the vacant Na2 site along path3 and then the inserted sodium atom sequentially moves to the vacant Na1 site along path2 (sequential mode). It is worth noting that both Na2-Na3-Na2 (path1) and Na2-Na1-Na2 (path2+path3) diffusion paths are two-dimensional in the $ac$ (or $bc$) plane. Fig.~\ref{fig_mig}(b) shows the calculated activation energies for these sodium migrations. Along path1, it was calculated to be $\sim$1.06 eV, which is significantly higher than the one along path2+path3 in the concurrent mode, $\sim$0.72 eV. Moreover, the length of the path1 channel ($\sim$14.4 \AA) is larger than those along path2+path3 ($\sim$11.7 \AA~in concurrent mode and $\sim$12.4 \AA~in sequential mode), so that we can reject the migration along path1. Meanwhile, it turns out that, although we let two sodium atoms move concurrently with a set of the NEB images accordingly, they actually moved in the sequential mode. We therefore regard the {\it two-dimensional pathway of path3+path2 in the sequential mode} as the actual channel for insertion-desertion of sodium atom. In this case, the activation energy for the preceding migration of the host sodium atom along path3 was found to be $\sim$0.56 eV, arriving at the final state where the two Na2 sites are occupied and the Na1 site is vacant, and that for the following migration of the inserted sodium atom along path2 was $\sim$0.16 eV. These values are within the range between 0.2 and 0.6 eV proper for SIB cathode with a fast rate~\cite{VanderVen,Ong}.

We then calculate the activation barriers for sodium migrations along path3 and path2 at different concentrations of inserted sodium atom. In the case of \ce{Na_{0.75}Fe(SO4)2} (overcharged state) the host sodium atom moves to the neighboring empty Na1 site passing through the unoccupied Na2 site, for which the activation energies were calculated to be $\sim$0.65 eV along path3 and 0.25 eV along path2. Along path3, we found the lowest value of 0.56 eV for the case of $x=1.75$, the middle value of 0.61 eV for $x=1.25$, and the highest value of 0.68 eV for $x=1.5$. The reasons for the lowest and highest values might be due to the large Na slab spacing for the case of $x=1.75$ and the weaker interaction between ions for the case of $x=1.25$. Although the absolute energy heights from the final states along path2 has similar tendency like 0.72, 0.78, and 0.82 eV for $x=1.75, 1.25$, and 1.5, the activation energies were more or less similar like 0.16, 0.17, and 0.14 eV, which are the energy difference between the transition and initial states.

At this point, it is worthwhile to compare our calculated activation barriers with other calculation results for different polyanionic compounds. For the case of \ce{Na2Fe2(SO4)3} calculated by Barpanda et al.~\cite{Barpanda14}, the lowest value in our work is comparable to 0.55 eV for the Na2 channel and 0.54 eV between Na1 and Na3 sites in that compound. Moreover, the authors in that work suggested that the Na1 ion can be extracted through the Na3 sites, being similar mechanism with our work. According to the calculation by Kim et al.~\cite{Kim_NaFePO4}, Na$_{1-x}$FePO$_4$ ($x=0$) has an unrealistic diffusivity of Na ion in the crystalline marcite phase due to a high activation energy of $\sim$1.46 eV, but in amorphous phase it can be promising due to a reasonable value of $\sim$0.73 eV, being similar to the absolute energy height in our work.

\begin{figure}[!t]
\begin{center}
\includegraphics[clip=true,scale=0.53]{fig3abc.eps} \\
\includegraphics[clip=true,scale=0.117]{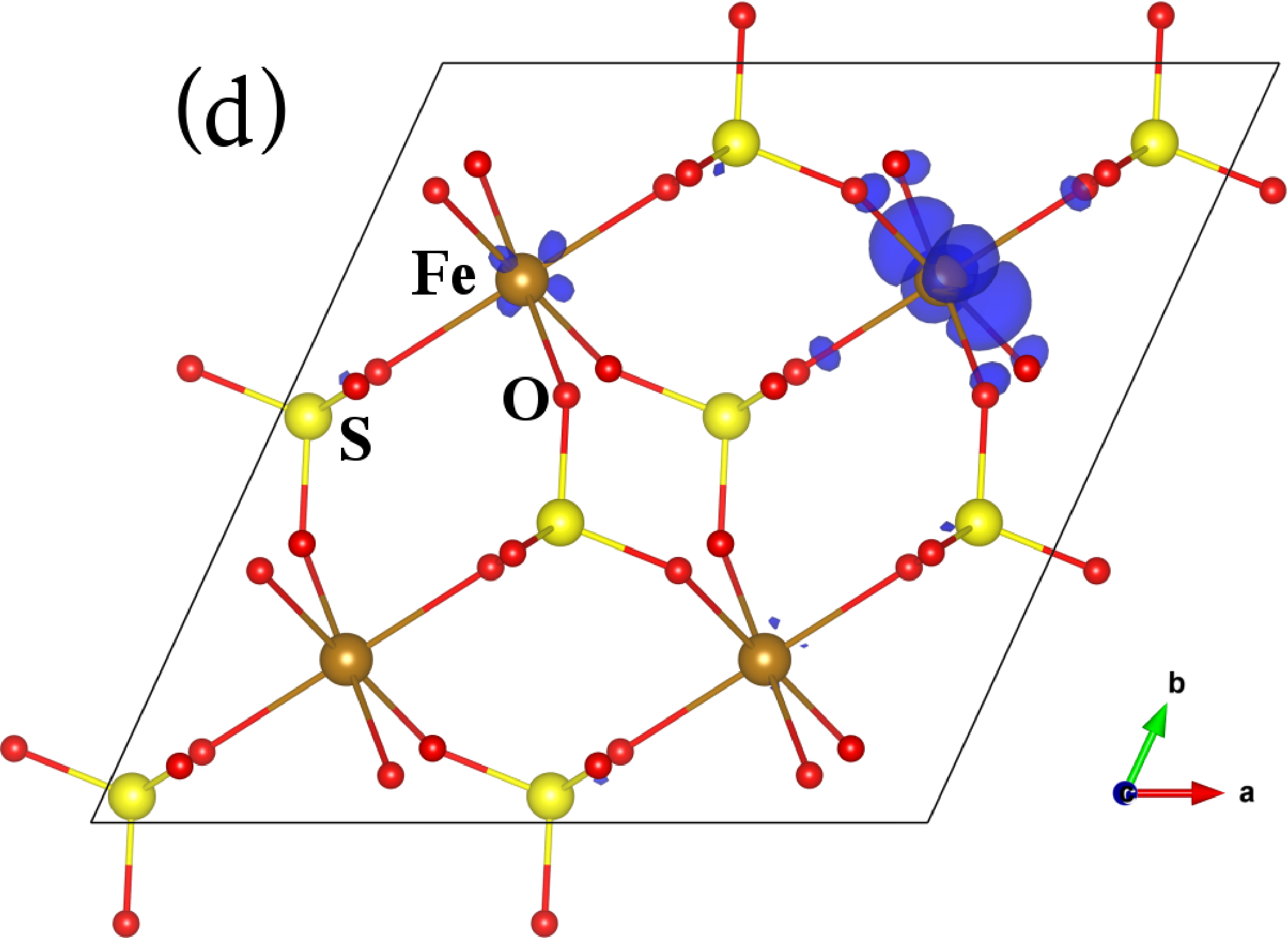}
\includegraphics[clip=true,scale=0.117]{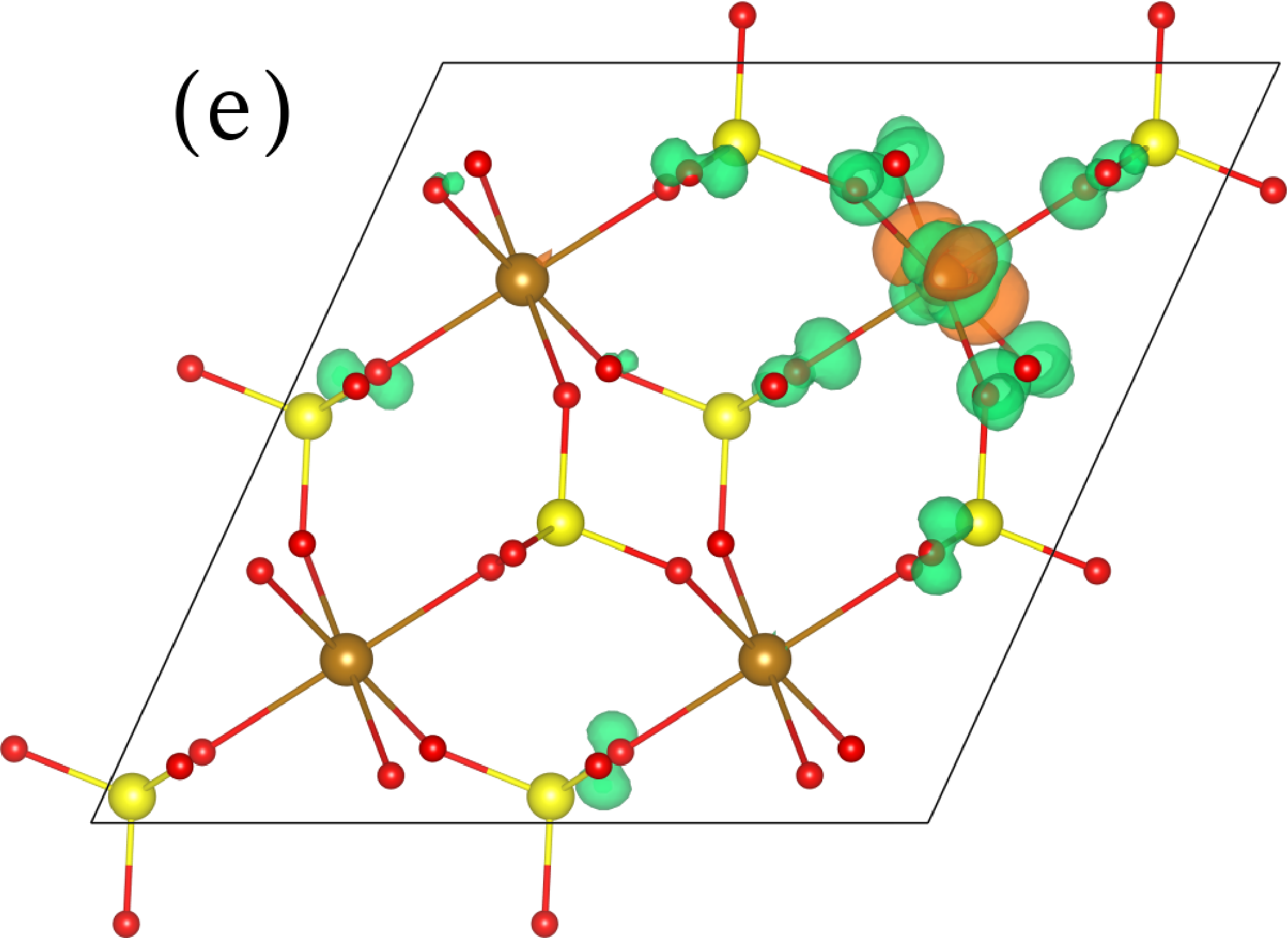}
\caption{\label{fig_dos}(color online) PDOS of (a) \ce{NaFe(SO4)2}, (b) \ce{Na_{1.25}Fe(SO4)2}, and (c) electron-injected \ce{NaFe(SO4)2}. Green arrows indicate the spin direction of electron. VBMs are set to be zero and shown by dashed line. Isosurfaces of (d) electronic charge density difference (value = 0.01 $\vert e\vert$/Bohr$^3$) in \ce{NaFe(SO4)2} between neutral and electron-injected states, and of (e) integrated local density of states of electron-injected \ce{NaFe(SO4)2} in the energy range of ($-0.5$, 0.7) eV.}
\end{center}
\end{figure}
Finally, we resolve the issue of electronic transfer, which is even more important as this compound is highly insulating material as discussed below. To get an insightful understanding of the electron transfer, we looked deep into the band structures and projected density of states (PDOS) of Na$_x$Fe(SO$_4$)$_2$ at different concentrations $x$. Regarding the band structures, we obtained a band gap of 1.7 eV at $x=1.0$, as an energy difference between the conduction band minimum (CBM) and the valence band maximum (VBM). When increasing the amount of inserted sodium atom, the band gaps gradually decrease to 1.1 ($x=1.25$), 0.5 ($x=1.5$), and 0.02 eV ($x=1.75$). It should be stressed that there is little cross between the bands with narrow bandwidth, indicating a localization of electronic state and a difficulty of electron transition between VBM and CBM. For the case of $x=1.75$, nevertheless, we persist it metallic, considering that its band gap is smaller than the energy of $k_BT\approx0.03$ eV at $T=300$ K. Through the analysis of PDOS, we revealed that Fe $3d$ and O $2p$ states predominantly govern the bands below VBM and over CBM, while observing some overlap between them, as in other transition metal oxides~\cite{Aydinol,MarianettiP,Hinuma}. The occupied and empty states of Na atom reside far from the band gap region, indicating a fully ionization of sodium. Upon insertion of sodium, an excess electron is released by ionization of sodium and moves to Fe $3d$ and O $2p$ states around CBM and VBM, forming impurity-type bands and reducing the band gaps as discussed above. Such characteristics of bands indicates that Na$_x$Fe(SO$_4$)$_2$ compounds at low compostion of inserted sodium are a band insulator and thus the mechanism of metallic electron conduction can not be applied.

Here we suggest a polaron hoping for the electronic transfer in this compound. When an electron is injected into the electrically neutral compound, the Jahn-Teller distortion of iron-oxygen octahedron is induced with a formation of polarizing field around the Fe atom. The electron can be trapped by this local lattice distortion and form a small polaron, which is the quasiparticle formed by the electron and its self-induced lattice distortion~\cite{Maxisch,McKenna,Kong,Longo,Sai}. In order to check the possibility of polaron formation, we compared the electronic properties of electron-injected \ce{NaFe(SO4)2} with those of \ce{Na_{1.25}Fe(SO4)2} as well as neutral \ce{NaFe(SO4)2}, as can be seen in Fig.~\ref{fig_dos}(a)-(c). Two bands with majority spin appear below VBM and one band with minority spin above CBM upon injection of an electron as evidencing by PDOS in Fig.~\ref{fig_dos}(c). Interestingly, electron-injected \ce{NaFe(SO4)2} is identical with \ce{Na_{1.25}Fe(SO4)2}, indicating that the inserted sodium atom is in full oxidation state as \ce{Na+} and the change of electronic structure upon insertion of sodium is due to the electron from it. In Fig.~\ref{fig_dos}(d) we plot the charge density difference between the neutral and the electron-injected \ce{NaFe(SO4)2}, showing {\it the localization of electron around one Fe atom and thus the formation of small polaron}~\cite{Maxisch}. Through the L\"{o}wdin charge analysis, the polaron forming Fe atom is confirmed to have 0.23 more electron than other Fe atoms. Fig.~\ref{fig_dos}(e) shows the integrated local density of states corresponding to the polaron states around Fe and O atoms.

We estimated a stability of the polaron by its self-trapping energy, which is a difference between the ionization energies of the localized electron state, $I_\text{loc}=E_\text{loc}^+-E_\text{ref}^0$, and the delocalized electron, $I_\text{deloc}=E_\text{deloc}^+-E_\text{ref}^0$, where $E^+$ and $E_\text{ref}^0$ refer to the total energy of the charged and the neutral state~\cite{Sai}. We found $I_\text{loc}\approx 3.72$ eV and $I_\text{deloc}\approx 4.07$ eV, obtaining the negative self-trapping energy ($E_\text{st}=I_\text{loc}-I_\text{deloc}$) of $-0.35$ eV, which means that the electron polaron is stable in this compound. The electron transfer is occurred by the polaron hoping from \ce{Fe_A^{2+}Fe_B^{3+}} to \ce{Fe_A^{3+}Fe_B^{2+}}. The activation energy for this polaron hoping was calculated to be $\sim$0.10 eV by using the linear interpolation approach~\cite{Maxisch,McKenna}, implying that the polaron is very mobile.

Let us compare with the conventional cathodes of LIB such as Li$_x$CoO$_2$ and Li$_x$FePO$_4$ ($0\leq x\leq 1$). On the contrary to Na$_x$Fe(SO$_4$)$_2$ ($1\leq x\leq 2$), Li$_x$CoO$_2$ is a band insulator at high Li composition and becomes metallic as decreasing Li composition~\cite{MarianettiN,Longo}, while Li$_x$FePO$_4$ has low intrinsic electron conductivity at both low and high Li composition~\cite{Maxisch}. Thus, the hole polaron (hole and electron polaron) formation after an electron removal (electron removal and injection) and the polaron hoping with the activation energy of $\sim$0.21 eV (0.215 and 0.175 eV) were confirmed for Li$_x$CoO$_2$ (Li$_x$FePO$_4$)~\cite{Longo}(\cite{Maxisch}).

In conclusion, we have studied the electrochemical properties, ion diffusion, and electron transfer of eldfellite Na$_x$Fe(SO$_4$)$_2$ using DFT+$U$ method. Our calculations reproduce the electrode potential in good agreement with the experiment, giving the small relative volume change. We suggest that the inserted sodium atom diffuses along the two-dimensional pathway accompanying with preceding movement of the host sodium atom with the reasonable activation energy for fast insertion and extraction. We also elucidate the electronic properties which indicate band insulating of the compound, and suggest the electron polaron formation around one iron atom and hoping between neighboring iron atoms. Our study may contribute to the understanding of possible mechanism of SIB operation based on eldfellite, and may open a new way for developing innovative SIB cathodes.

This work is supported as part of the fundamental research project ``Design of Innovative Functional Materials for Energy and Environmental Application'' (no. 2016-20) funded by the State Committee of Science and Technology, DPR Korea. Computation was done on the HP Blade System C7000 (HP BL460c) that is owned by Faculty of Materials Science, Kim Il Sung University.

%\bibliographystyle{apsrev}
%\bibliography{Reference}

\end{document}

\title{Supplemental Material -- Nature of Ionic Diffusion and Electronic Transfer in Eldfellite Na$_x$Fe(SO$_4$)$_2$}

\author{Chol-Jun Yu$^{1*}$, Song-Hyok Choe$^1$, Gum-Chol Ri$^1$, Sung-Chol Kim$^1$, Hyok-Su Ryo$^2$, and Yong-Jin Kim$^2$}
\affiliation{$^1$Department of Computational Materials Design (CMD), Faculty of Materials Science, and\\
$^2$Faculty of Physics, Kim Il Sung University, Ryongnam-Dong, Taesong District, Pyongyang, DPR Korea}
%\date{}
\maketitle

\section*{Crystal structure}
\begin{figure}[!th]
\centering
\includegraphics[clip=true,scale=0.33]{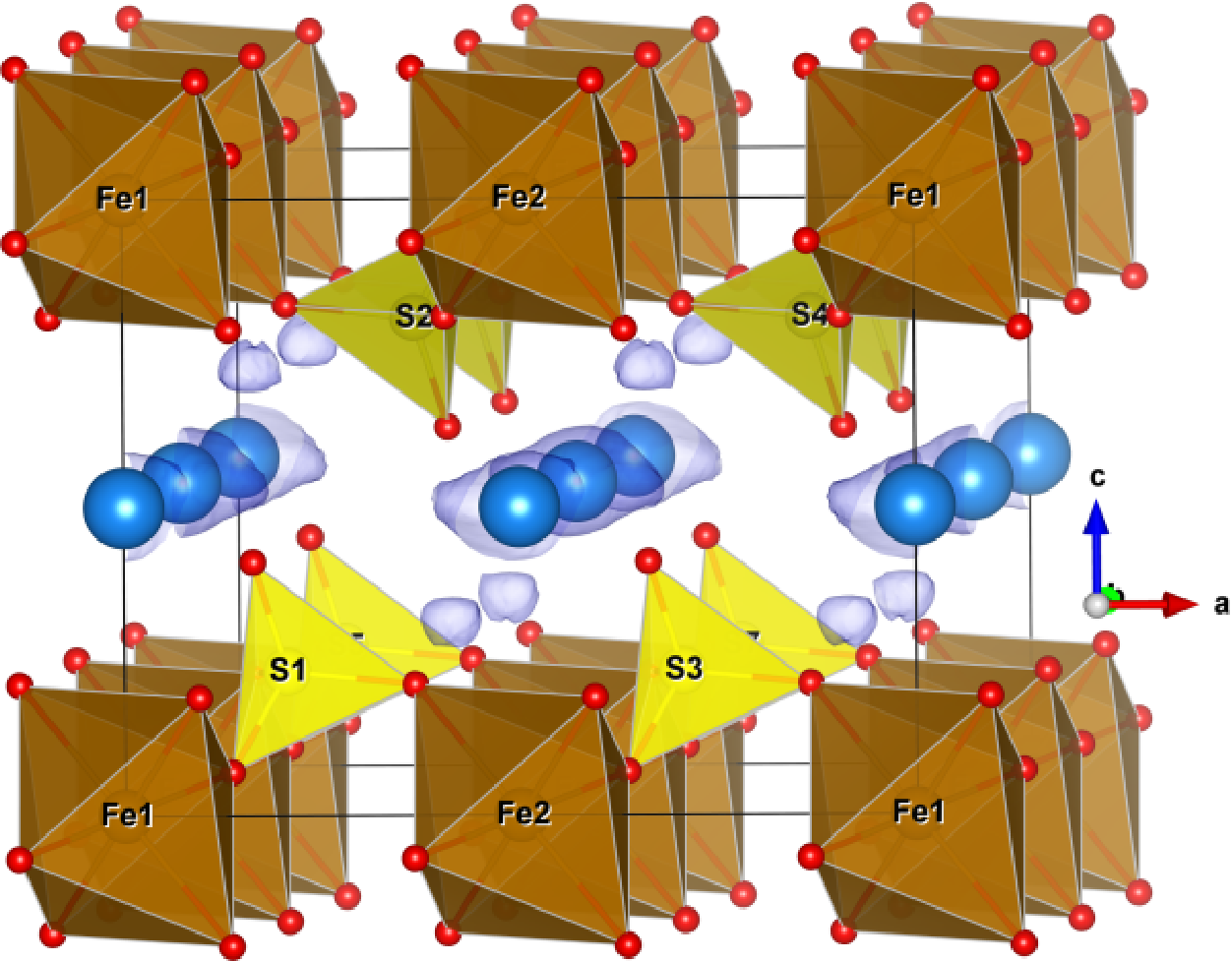}
\includegraphics[clip=true,scale=0.15]{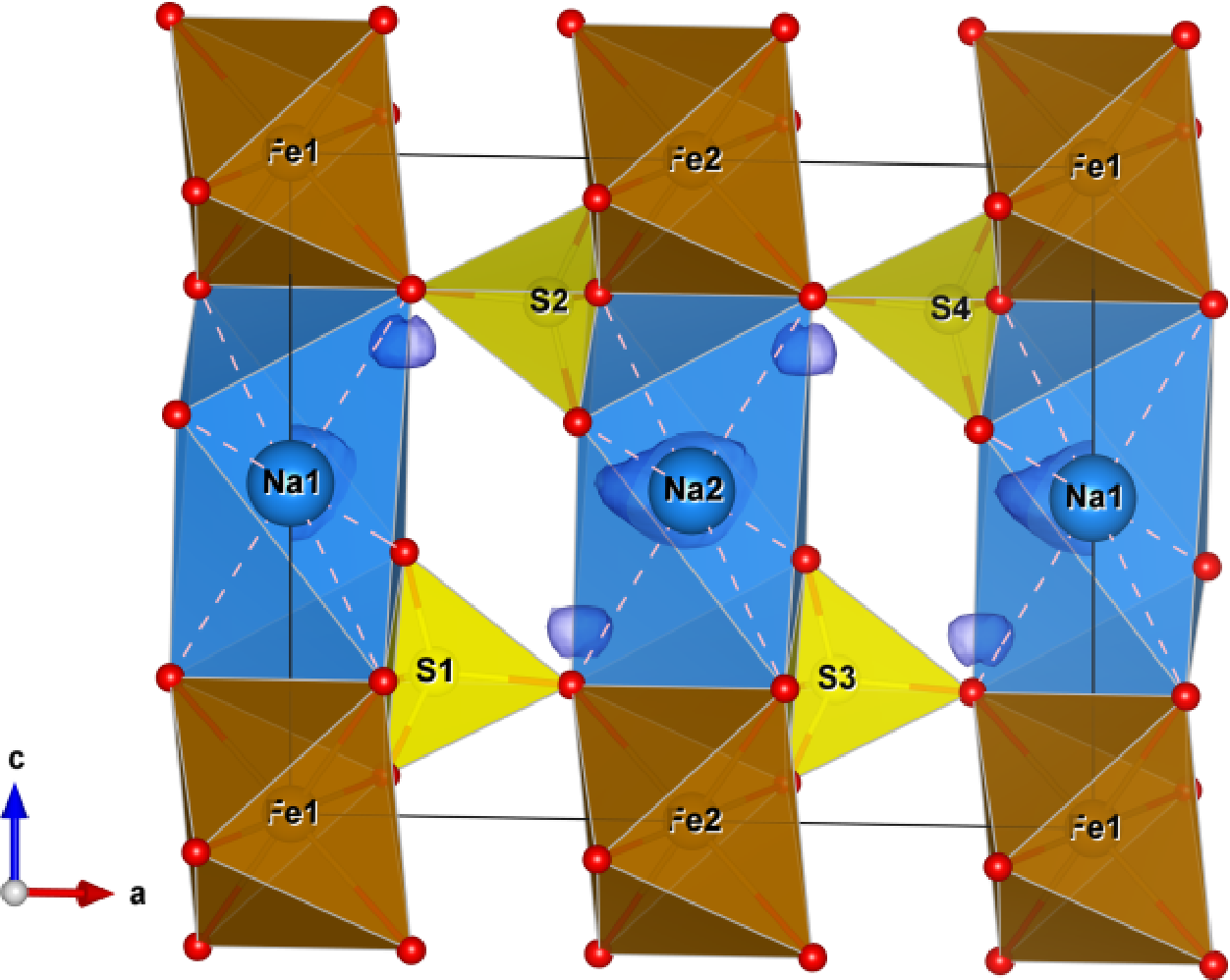}
\includegraphics[clip=true,scale=0.23]{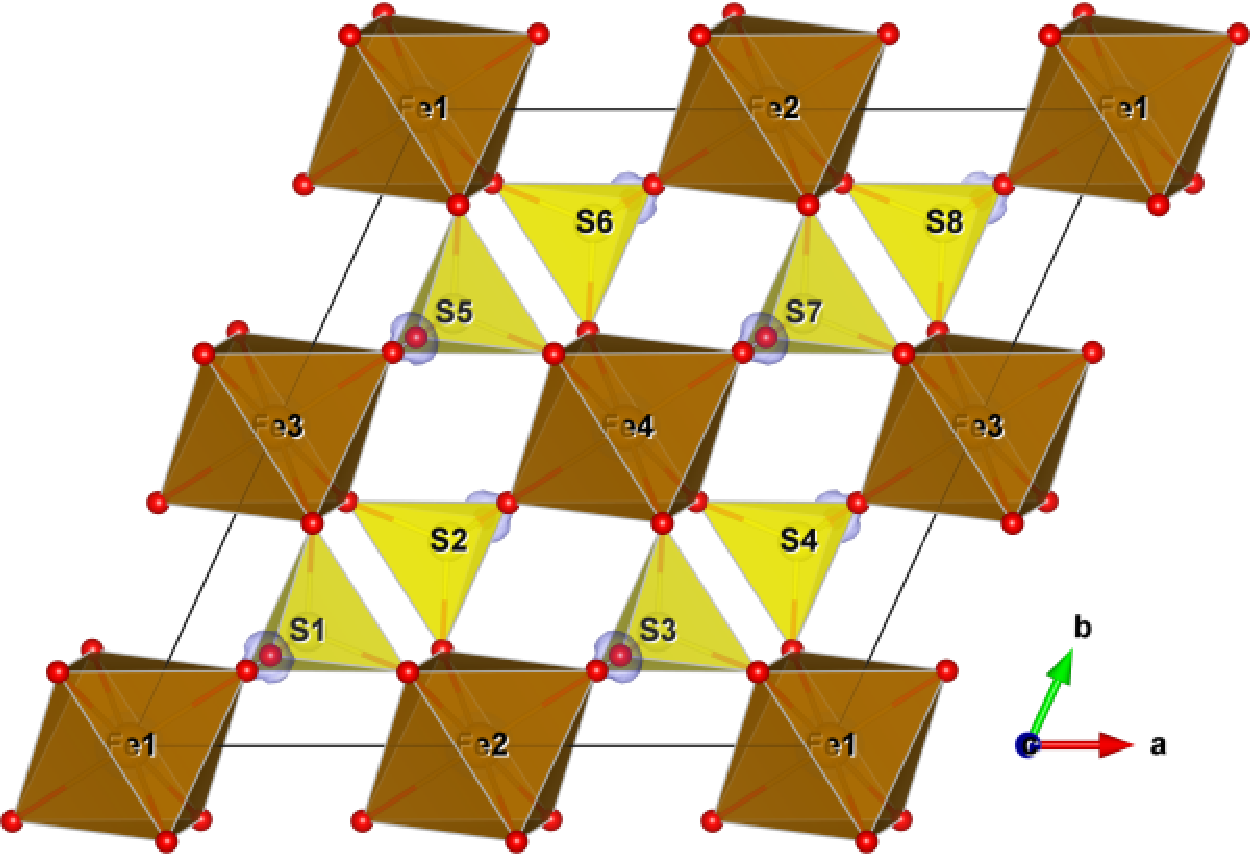}
\caption{\label{fig_str}Polyhedral view of $(2\times2\times1)$ supercells containing 4 formula units for bulk eldfellite \ce{NaFe(SO4)2}. Sodium and oxygen atoms are represented by big blue and small red balls. For clear view of layered structure, distorted sodium octahedra is also shown in the bottom-right panel. Isosurface map of $\Delta$BVS (value = 0.3) is also shown to indicate the possible positions of inserted sodium atoms.}
\end{figure}

Eldfellite \ce{NaFe(SO4)2} crystallizes in a layered structure with monoclinic space group $C2/m$, in which the layers are composed of interconnected \ce{FeO6} octahedra and distorted \ce{NaO6} octahedra in the $a$-$c$ plane or equivalently $b$-$c$ plane. These planes are bridged by \ce{SO4} tetrahedra that leave interplanar space for 2D \ce{Na+} or \ce{Li+} guest-ion diffusion~\cite{Zunic,ReynaudPhd,Singh}. Fig.~\ref{fig_str} shows the polyhedral view of its optimized crystal structure.

To investigate the intercalation of Na ion, we first built the primitive unit cell that contains one formula unit (1f.u., 12 atoms) and constructed the (2$\times$2$\times$1) supercell (4f.u., 48 atoms). For structural optimization, all atoms and cell parameters were fully relaxed so that Jahn-Teller distortions were allowed where the Fe ions are Jahn-Teller active.

\section*{Computational details}
\begin{figure}[!th]
\includegraphics[clip=true,scale=0.48]{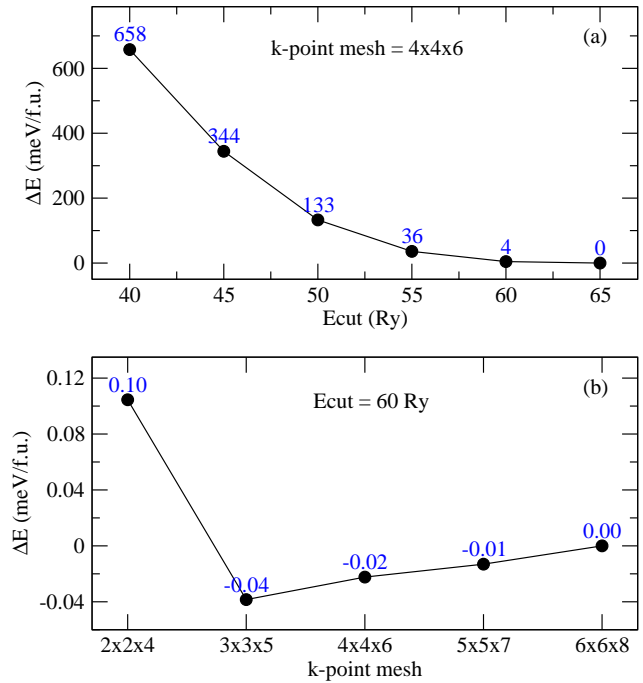} 
\caption{\label{fig_conv}DFT total energy convergence as increasing (a) plane wave cutoff energy and (b) $k$-point mesh.}
\end{figure}

We performed density functional theory calculations using the pseudopotential plane-wave code Quantum ESPRESSO (version 5.3)~\cite{QE}. Plane wave cutoff energy and $k$-point mesh were tested for the (2$\times$2$\times$1) supercell of \ce{NaFe(SO4)2}, as presented in Fig.~\ref{fig_conv}. A cutoff energy of 60 Ry was found to be sufficient for total energy convergence of 5 meV/f.u. as shown in Fig.~\ref{fig_conv}. We used a $k$-point mesh of 2$\times$2$\times$4 for structural relaxations, while denser $k$-point mesh of 6$\times$6$\times$8 for electronic structure calculations. Self-consistent convergence thresholds for the total energy and force were set to $10^{-9}$ Ry and 5$\times10^{-4}$ Ry/Bohr. We applied Methfessel-Paxton smearing approach with a 0.2 Ry gaussian spreading for structural optimization, while tetrahedra approach for electronic structure calculations.

All the calculations were spin polarized with starting spin polarization of 1 for all Fe atoms (ferromagnetic: FM) or $-1$ for half Fe atoms (antiferromagnetic: AFM) and 0 for other atoms. The total energy difference between AFM and FM states of Na$_x$Fe(SO$_4$)$_2$ indicates that AFM phases are slightly more stable than FM phases in $x=0.75$, 1.0, 1.25, 1.5 and 1.75. In Table~\ref{tab_lattice}, we show the calculated lattice parameters of (2$\times$2$\times$1) supercells for \ce{Na_{\it x}Fe(SO4)2} ($0.75\leq x\leq1.75$) in FM state and the total energy differences from AFM state.
\begin{table}[!th]
\caption{\label{tab_lattice} Lattice constants and angles, and volumes (\AA$^3$) of (2$\times$2$\times$1) supercells of \ce{NaFe(SO4)2} in FM state, and the total energy difference from AFM state, calculated by PBE+$U$ method with $U=4.0$ eV.}
\begin{tabular}{c|ccccccc}
\hline
$x$ & $a$ & $c$ & $\alpha$ & $\gamma$ & Volume & $\Delta V$ & $\Delta E_\text{tot}$\\
    & (\AA) & (\AA) & ($^\circ$) & ($^\circ$) & (\AA$^3$) & (\%) & (meV/cell) \\
\hline
0.75 & 9.887 & 7.331 & 91.71 & 64.88 & 648.347 & 1.50 &  2.18 \\
1.00 & 9.833 & 7.284 & 91.69 & 64.99 & 638.794 &      &  4.50 \\
1.25 & 9.887 & 7.312 & 91.70 & 64.88 & 646.771 & 1.25 &  1.68 \\
1.50 & 9.955 & 7.285 & 91.71 & 64.73 & 651.017 & 1.91 &  3.03 \\
1.75 & 9.999 & 7.245 & 91.72 & 64.63 & 654.617 & 2.48 &  0.62 \\
\hline
\end{tabular}
\end{table}

To estimate the possible positions of inserted \ce{Na+} ions and the \ce{Na+} ion diffusion paths, bond valence sum (BVS) method was applied, where BVS can be calculated as follows,
\begin{equation}
B(\mathbi{r})=\sum_i \exp\left[\frac{R_0-R_i(\mathbi{r})}{b}\right]
\end{equation}
Here, $R_i(\mathbi{r})=|\mathbi{r}-\mathbi{R}_i|$ ($i$ only for oxygens), and $R_0=1.803$ \AA~and $b=0.370$ \AA~for the Na$-$O bond~\cite{Brown_BVS}. The values of $B(\mathbi{r})$ at the positions of host sodium atoms were evaluated to be almost 1 and the difference of $B(\mathbi{r})$ from this value were calculated for whole space with a grid resolution of 0.1 \AA.

The electrode voltage can be derived from DFT calculations with good accuracy~\cite{Aydinol,Wolverton}. Upon sodium intercalation into a cathode host represented by the equation,
\begin{equation}
\ce{Na_{\it x_i}Fe(SO4)2}+(x_j-x_i)\ce{Na_{bcc}}\rightarrow \ce{Na_{\it x_j}Fe(SO4)2}
\end{equation}
where $x_i$ and $x_j$ are the limits of the intercalation reaction and \ce{Na_{bcc}} is a sodium metal anode in bcc phase, the average equilibrium cell voltage was approximately predicted by the DFT total energy change per intercalated \ce{Na+} ion as follows,
\begin{equation}
V=-\frac{E_\ce{Na_{\it x_j}Fe(SO4)2} - \left[E_\ce{Na_{\it x_i}Fe(SO4)2} + (x_j-x_i)E_\ce{Na_{bcc}}\right]}{(x_j-x_i)e}
\end{equation}
where $E_\ce{Na_{\it x_j}Fe(SO4)2}$ is the DFT total energy of the (2$\times$2$\times$1) supercell for $\ce{Na_{\it x_j}Fe(SO4)2}$, $E_\ce{Na_{bcc}}$ is the energy per atom of Na in the bcc crystal, and $e$ is an electronic charge. Binding energy of intercalated \ce{Na+} ion, $E_b$, can be calculated as follows,
\begin{equation}
E_b=E_\ce{Na_{\it x}Fe(SO4)2} - \left[E_\ce{NaFe(SO4)2} + (x-1)E_\ce{Na_{gas}}\right]
\end{equation}
where $E_\ce{Na_{gas}}$ is the DFT total energy of isolated Na atom in its gaseous state.

\section*{NEB calculation}
\begin{figure}[!th]
\includegraphics[clip=true,scale=0.28]{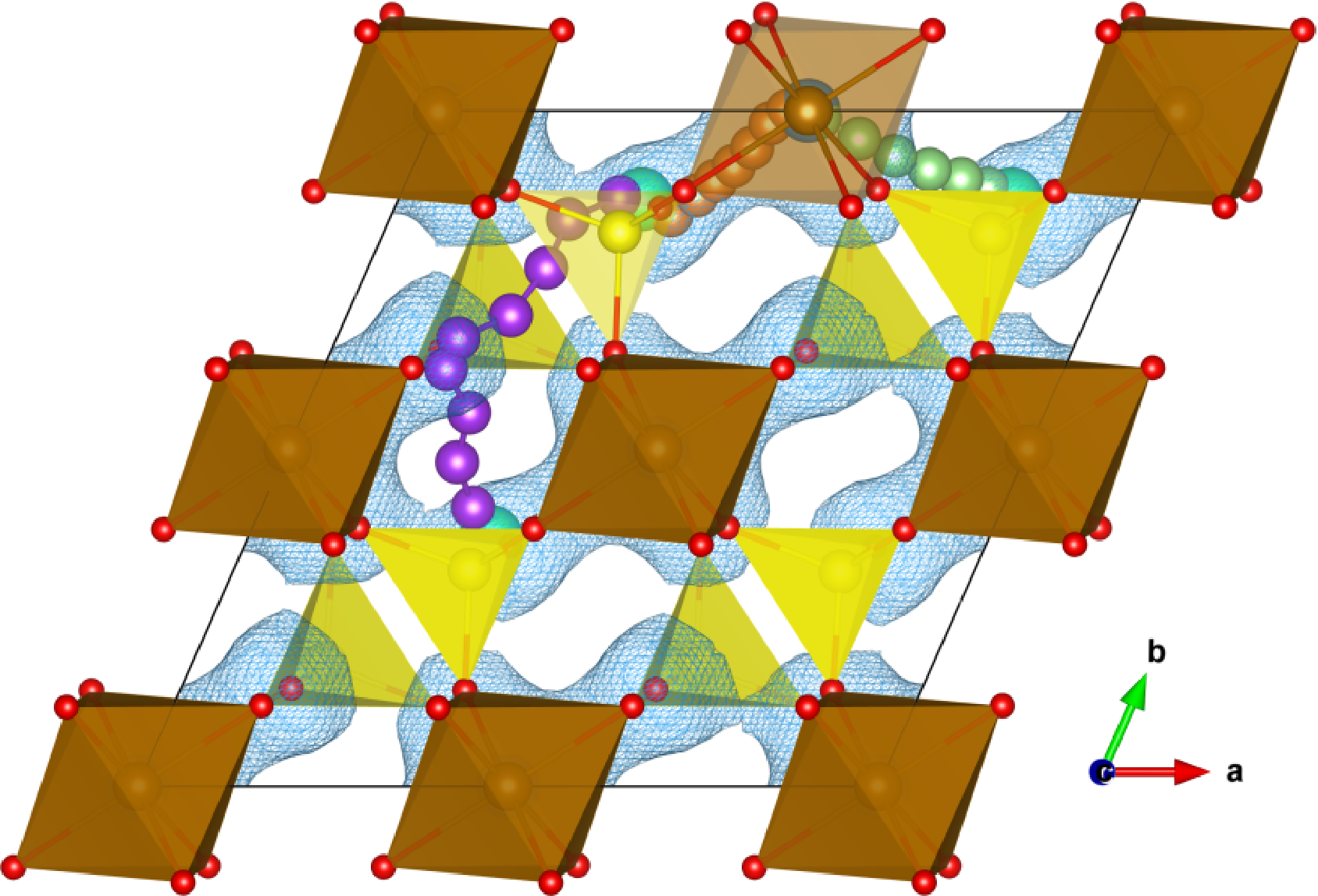}
\caption{\label{fig_neb}Pathways of Na2-Na3-Na2 channel (path1) and Na2-Na1-Na2 channel (path3+path2) in $ab$ plane with an isosurface of $\Delta$BVS (value = [0.3, 1.7]).}
\end{figure}

To calculate the migration barriers for sodium ion diffusion, we applied the climbing image nudged elastic band (NEB) method~\cite{NEB} to \ce{Na_{\it x}Fe(SO4)2} with $x=1.0$, 1.25, 1.5, and 1.75, using $(2\times2\times1)$ supercells. The supercell dimensions were fixed at the optimized supercell size during the NEB runs, while all the atoms were relaxed with the convergence criteria for the forces of 0.02 eV/\AA. Two pathways are possible for the inserted sodium atom migration; (i) path1: Na2-Na3-Na2 channel, and (ii) path2+path3: Na2-Na1-Na2 channel. In the latter case, since the Na1 site is occupied by the host sodium atom, two modes are possible; (1) concurrent move: the inserted and host sodium atoms move concurrently, and (2) the host sodium atom moves first to the Na2 site and then the inserted sodium atom moves to the vacant Na1 site. Fig.~\ref{fig_neb} shows the pathways for migration of the inserted sodium atoms. To allow the concurrent move of two sodium atoms along path2+path3, two sets of NEB images along path2 and path3 are settled. The number of images adopted in this work were 11 and 7 according to the length of pathway.

\section*{Electronic structure}
\begin{figure}[!th]
\includegraphics[clip=true,scale=0.3]{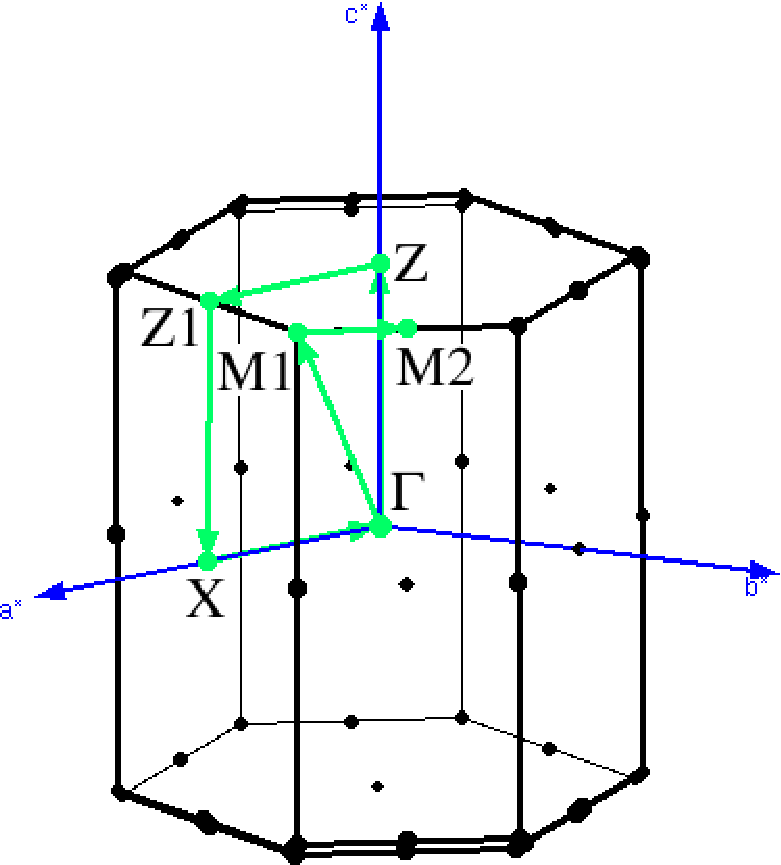}
\caption{\label{fig_bz}First Brillouin-zone in reciprocal space.}
\end{figure}

We first perform self-consistent field (SCF) calculation of optimized $(2\times2\times1)$ supercells for the compounds with the $k$-points of $6\times6\times8$ and the tetrahedra option for occupations, and then non-SCF calculations for band structures with a $k$-path passing high symmetry points and lines shown in Fig.~\ref{fig_bz}.
\begin{figure*}[!th]
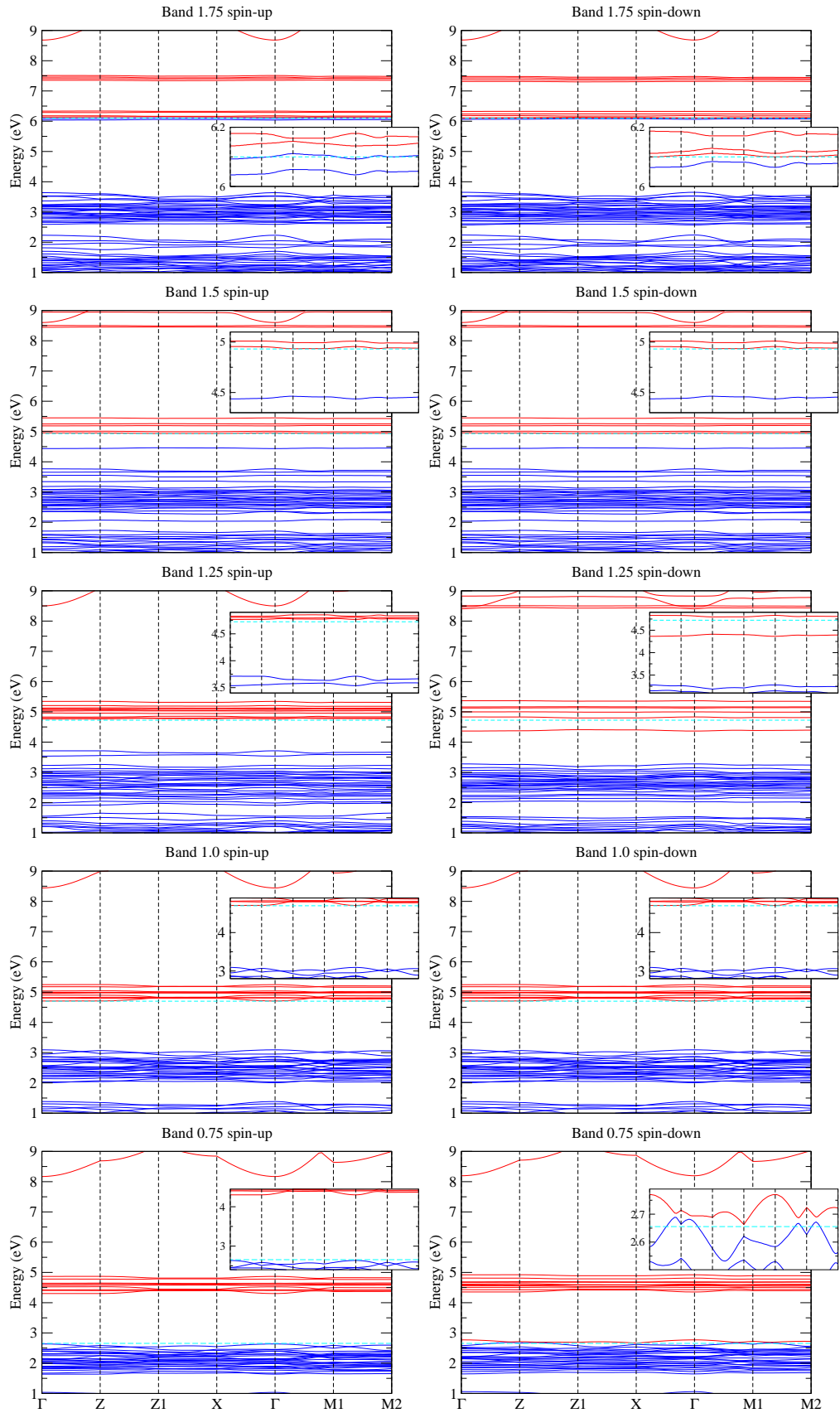

\begin{tabular}{cc}
\includegraphics[clip=true,scale=0.4]{figs5a.eps} & \includegraphics[clip=true,scale=0.4]{figs5b.eps} \\
\includegraphics[clip=true,scale=0.4]{figs5c.eps} & \includegraphics[clip=true,scale=0.4]{figs5d.eps} \\
\includegraphics[clip=true,scale=0.4]{figs5e.eps} & \includegraphics[clip=true,scale=0.4]{figs5f.eps} \\
\includegraphics[clip=true,scale=0.4]{figs5g.eps} & \includegraphics[clip=true,scale=0.4]{figs5h.eps} \\
\includegraphics[clip=true,scale=0.4]{figs5i.eps} & \includegraphics[clip=true,scale=0.4]{figs5j.eps} \\
\end{tabular}
\caption{\label{fig_band}Electronic band structures of Na$_x$Fe(SO$_4$)$_2$ at $x=0.75$, 1.0, 1.25, 1.5, and 1.75. Blue and red lines show valence and conduction bands.}
\end{figure*}
In Fig.~\ref{fig_band}, we show the electronic band structures of Na$_x$Fe(SO$_4$)$_2$ with $(2\times2\times1)$ supercells by using PBE+$U$ method with $U=4.0$ eV. 

\begin{figure}[!th]
\includegraphics[clip=true,scale=0.23]{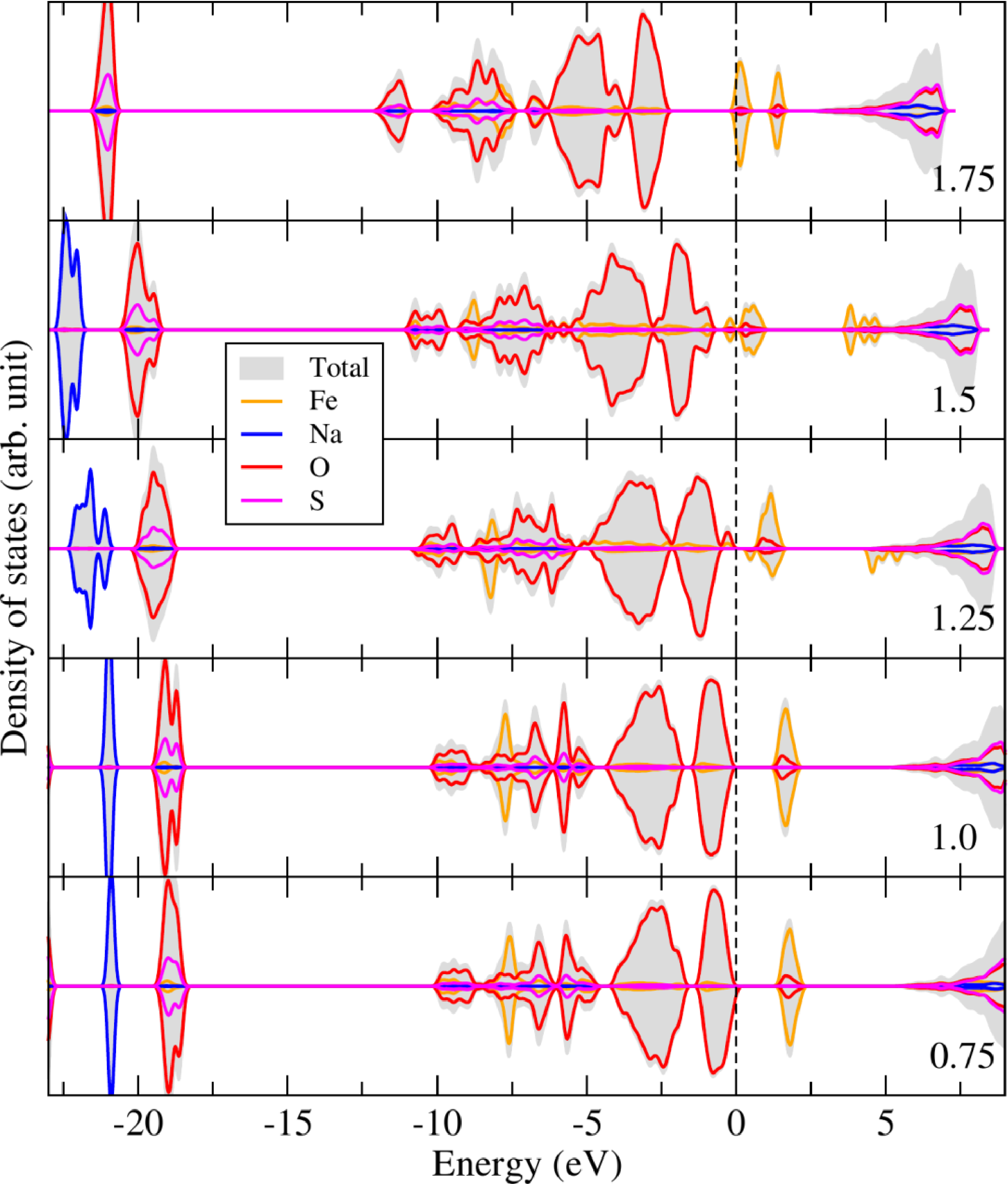}
\caption{\label{fig_fulldos}Density of states of Na$_x$Fe(SO$_4$)$_2$ at $x=0.75$, 1.0, 1.25, 1.5, and 1.75. Valence band maximums are set to be zero.}
\end{figure}
\begin{figure}[!th]
\includegraphics[clip=true,scale=0.5]{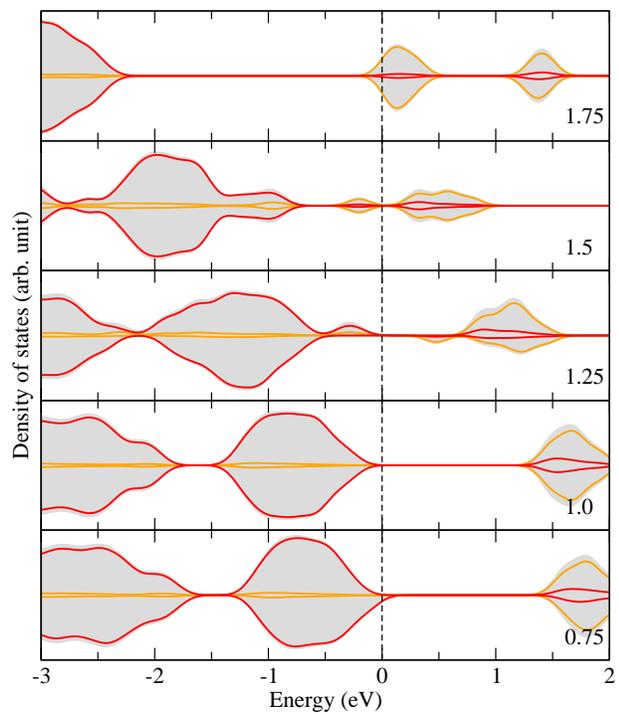}
\caption{\label{fig_dosdet}Density of states of Na$_x$Fe(SO$_4$)$_2$  at $x=0.75$, 1.0, 1.25, 1.5, and 1.75 in the vicinity of valence band maximum.}
\end{figure}
We calculate the partial density of states of Na$_x$Fe(SO$_4$)$_2$ at $x=0.75$, 1.0, 1.25, 1.5, and 1.75, as shown in Figs.~\ref{fig_fulldos} and~\ref{fig_dosdet}.

%\bibliographystyle{apsrev}
%\bibliographystyle{plain}
%\bibliography{Reference}